\documentclass[aps,prd,twocolumn,groupedaddress,showpacs,showkeys]{revtex4}

\usepackage{graphicx}

\newcommand{\nn}{\nonumber\\}

\def\be{\begin{equation}}
\def\ee{\end{equation}}
\def\bea{\begin{eqnarray}}
\def\eea{\end{eqnarray}}

\begin{document}


\title{Small-$x$ QCD evolution of $2 n$ Wilson line correlator: the weak field limit}
\author{Alejandro Ayala$^{1,4}$, Erike R. Cazaroto$^{2,3}$, Luis Alberto Hern\'andez$^{4}$, Jamal Jalilian-Marian$^3$, Maria Elena Tejeda-Yeomans$^5$}
\affiliation{$^1$Instituto de Ciencias Nucleares, Universidad Nacional Aut\'onoma de M\'exico, 
Apartado Postal 70-543 M\'exico Distrito Federal 04510, M\'exico\\
$^2$Instituto de F\'{\i}sica, Universidade de S\~{a}o Paulo, 
C.P. 66318,  CEP 05315-970, S\~{a}o Paulo, SP, Brazil\\
$^3$Department of Natural Sciences, Baruch College, CUNY,
17 Lexington Avenue, New York, NY 10010, USA\\
$^4$Centre for Theoretical and Mathematical Physics, and Department 
of Physics, University of Cape Town, Rondebosch 7700, South Africa\\
$^5$Departamento de F\1sica, Universidad de Sonora, Boulevard
Luis Encinas J. y Rosales, Colonia Centro, Hermosillo, Sonora 83000, Mexico.
}

\begin{abstract}
\noindent We write down explicit expressions for the $x$-evolution (equivalent to energy or rapidity evolution) 
of $2 n$ ($n = 1, 2, \cdots$) Wilson lines using the JIMWLK equation and the Color Glass Condensate formalism. We investigate the equation in the weak gluon field limit (linear regime) by expanding the Wilson lines in powers of the gluon field and show that it reduces to the BJKP equation describing the evolution of a state of $2 n$ Reggeized gluons with energy. We also make available for download a {\it Mathematica} program which provides this expression for any value of $n$. 

\end{abstract}



\maketitle

\section{Introduction} 
Gluon saturation is expected to play a dominant role in high energy scattering processes where there is at least one hadron/nucleus in the initial state of the collision~\cite{cgc-reviews}. At high center of mass collision energy and as long as the transverse momenta involved in the process are not too large, the biggest contribution to the cross section comes from the small $x_{Bj}$ kinematics, where the wave function of the incoming hadron/nucleus consists mostly of gluons. The Color Glass Condensate (CGC) formalism is an effective theory approach to QCD at high energy that includes two effects which are expected to be significant and are not contained in the leading twist pQCD approach to hadronic scattering: First, multiple scatterings (in the target frame) become important when the target hadron/nucleus becomes a dense system of gluons. Second, large logarithms of energy become more important than the not so large logarithms of transverse momentum. 

The inclusion of high gluon density effects, is accomplished by introducing a color current $J^a_\mu$ that represents the large $x$ degrees of freedom in the hadron/nucleus. The color charges $\rho^a$ generate a color field $A^a_\mu$ representing the small $x$ gluons in the target. In the original McLerran-Venugopalan model~\cite{mv} of a large nucleus the color charges are assumed to be random and are described by a local Gaussian profile in the two-dimensional transverse plane. It is also possible to include the longitudinal structure of the color sources~\cite{longitudinal}, as well as higher order terms (beyond Gaussian distributions) in the effective action~\cite{quartic} describing the large $x$ color charges. On the other hand, the inclusion of large logarithms of energy, is accomplished via a Wilsonian renormalization group equation, known as the JIMWLK equation~\cite{jimwlk} which describes the change in the distribution of the color charges as one goes to smaller $x$ (larger rapidities or energies). This equation can also be used to compute the $x$ evolution of any observable of theory. It is a non-closed functional equation which is equivalent to Balitsky hierarchy and reduces to the Balitsky-Kovchegov (BK) equation in the large $N_c$ and mean field limit~\cite{bk}. The most important aspect of the CGC formalism is the appearance of a dynamically generated semi-hard scale, called the saturation scale $Q_s$, which grows with decreasing $x$, increasing nucleon number $A$ and centrality. $Q_s$ can be much larger than $\Lambda_{QCD}$ which warrants a weak coupling (yet non-perturbative) approach to problems which have been traditionally thought to be outside the realm of weak coupling techniques. The analysis of multi-parton production and produced parton multiplicity distributions in hadronic/nuclear collisions, is the prime example of applicability of the CGC approach where leading twist pQCD is not applicable.  

The CGC formalism has been applied to DIS, proton-proton, proton (deuteron)-nucleus and nucleus-nucleus scattering processes at high energy colliders such as HERA, RHIC and the LHC. A very interesting question in nucleus-nucleus collisions of whether the produced partons thermalize (or become isotropic), can be investigated using the CGC formalism~\cite{cgc-thermal}. In addition, distributions of produced partons in high energy heavy-ion collisions can also be obtained, with some approximations, from the CGC formalism and used as initial conditions for hydrodynamic/transport description of the produced Quark-Gluon Plasma~\cite{cgc-hydro}. On the other hand, if one is solely interested in probing CGC and its properties, DIS offers the best environment since many theoretical complications  
can be avoided due to the color singlet nature of the virtual photon in the initial state. One can consider inclusive and diffractive structure functions, single and double inclusive particle production as well as diffractive particle 
production~\cite{cgc-dis}.
A complementary channel is particle production in high energy proton-nucleus (pA) collisions in the forward rapidity region at RHIC and the LHC. In the absence of a high energy electron-ion collider this is perhaps the best context in which to study CGC and gluon saturation. A typical particle production cross section in pA collisions involves correlations of multiple Wilson lines at different transverse coordinates. For instance, single inclusive photon and hadron production~\cite{cgc-single} involve two-point functions of Wilson lines (in fundamental or adjoint representation), the so called dipole, which also appears in DIS structure functions. Less inclusive processes such as di-hadron production~\cite{cgc-double} involve correlators of higher numbers of Wilson lines such as quadrupoles, sextuples, etc., with their energy dependence given by the JIMWLK equation~\cite{jimwlk}.

The energy dependence of the dipole, a $2$-point function of Wilson lines, has been extensively studied. At large $N_c$ and mean field it is governed by the BK equation, NLO corrections to which have been recently computed. Numerical solutions to the BK equation have been extensively used to investigate particle production in $pA$ collisions~\cite{rcBK}. On the other hand, higher point functions of Wilson lines, odderons, quadrupoles, etc., appear only in less inclusive processes and are much less studied than dipoles~\cite{cgc-multipole}. One exception is the kinematic region where the gluon field is weak so that Wilson lines can be expanded in powers of the gluon fields and only the first few terms are kept. This corresponds to investigating multi-particle production in the kinematic region where the produced particles have transverse momenta higher than the saturation scale in the problem. It has been shown that in this limit the evolution equation for the dipole is the well-known BFKL equation~\cite{bfkl}, whereas the evolution equation for the odderon and quadrupole reduces to the BJKP equation for the energy dependence of a state of $3$ or $4$ Reggeized gluons~\cite{cgc-odd, cgc-quad, bjkp} (see also \cite{simon} for a nice discussion of gluon Reggeization). It should be noted that this problem has also been investigated using a dual approach known as the KLWMIJ equation~\cite{klwmij}.

Our goal in this paper is to show, via an explicit calculation, that this reduction is in general true for the correlator of any $2\, n$ ($n = 1,2, 3, \cdots$) Wilson lines in the dilute limit, i.e., when it is justifiable to keep the lowest non-trivial term in the expansion of the Wilson line. This formally establishes the equivalence of the BJKP equation with the dilute limit of the JIMWLK equation. The work is organized as follows: In Sec.~\ref{II} we recall the basics of the JIMWLK evolution equation and derive an explicit equation for the evolution of the $2n$ Wilson line correlator. The automatic implementation of this procedure is performed using a {\it Mathematica}~\cite{math} program which returns the desired evolution equation after an integer value for $n$ is inserted~\cite{program}. We make the code available for download. In Sec.~\ref{III} we study the dilute or linear regime of this equation and give the general expression for the correlator of $2n$ Wilson lines in momentum space and show that it agrees with the BJKP equation~\cite{bjkp}. We finally summarize and conclude in Sec.~\ref{concl}. Explicit expressions for the evolution equation are given in the Appendix.

\section{The JIMWLK evolution equation for the $2n$ Wilson line correlator}\label{II}

The JIMWLK equation describes the $x$ (or energy) evolution of any observable $O$ in the CGC formalism. It reads  
\be
\frac{\partial < O >_Y}{\partial Y} \, = \, <H\, O>_Y \, 
\label{jimwlk-O}
\ee
where the Hamiltonian $H$ is given by 
\bea
       H &=& -\frac{1}{16 \pi ^3} \int d^2 x\, d^2 y \, d^2 z\nn
          &\times& M_{xyz} \left(1 + U_x^\dagger U_y - U_x^\dagger U_z - U_z^\dagger U_y  \right)^{ab}\nn 
          &\times&\frac{\delta}{\delta \alpha _x^a} \frac{\delta}{\delta \alpha _y^b},
\label{ham}
\eea
where $x,y,z$ are two-dimensional vectors on the transverse plane, $a$ and $b$ are color indices, and the {\it dipole} kernel $M_{xyz}$ is defined as
\be
M_{xyz} \equiv \frac{(x-y)^2}{(x-z)^2(z-y)^2}. 
\label{kernel}
\ee
In our notation the expression for the Wilson line $V$ is given by 
\be
V (x) \equiv e^{- i g \, \alpha^a (x)\, t^a},
\label{fundamental}
\ee
with $t^a$ a $SU (N)$ matrix in the fundamental representation. We use the notation where $U$ represents the analogous Wilson line in the adjoint representation. The gluon field $\alpha^a (x)$ is related to the color charge density $\rho^a (x)$ (in the covariant gauge) via
\be
\partial_\perp^2 \alpha^a (x) = - g \rho^a (x).
\ee

In this paper we are interested in the evolution equation for the $2 n$ Wilson line correlator $\hat{S}^{(2n)}$ 
(in the fundamental representation) defined as 
\bea
\hat{S}^{(2n)}_{\left(\prod _{k=1}^{2n} x_k\right)} = \frac{1}{N_c} Tr\left(V_{x_1} V_{x_2}^\dagger \, V_{x_3} V_{x_4}^\dagger \cdots V_{x_{2n-1}} V_{x_{2n}}^\dagger \right).
\label{2n_op}
\eea
In our notation, the upper index $\,2n\,$ of $\,\hat{S}\,$ represents the number of Wilson lines in this operator, and the lower index $\,\prod _{k=1}^{2n} x_k\,$ represents the coordinates of the Wilson lines, with $x_k$ being two-dimensional vectors on the transverse plane. Since the Wilson lines do not commute, the order in which the coordinates appear in $\,\hat{S}\,$ is important; it is exactly the order in which the Wilson lines appear inside the trace. For example, in Eq.~(\ref{2n_op}) it is understood that $\,\prod _{k=1}^{2n} x_k\,$ would be written out explicitly in increasing order in $\,x_i\,$: $\,x_1\,x_2\,x_3\,\cdots\,x_{2n}\,$. To be specific, when $\,n = 1\,$ the operator $\hat{S}^{(2n)}$ is the dipole operator, whereas for $\,n=2\,$ it is the quadrupole operator, and so on. 
Explicit evolution equations for the dipole ($2n=2$), quadrupole ($2n=4$) and sextuple ($2n=6$) have been derived \cite{cgc-multipole}. Here we derive and provide an explicit expression for the evolution of $\,\hat{S}^{(2n)}\,$ correlator of $2n$ Wilson lines for any $n$. 

The first step is to successively apply the functional double derivative on the operator $\,\hat{S}^{(2n)}\,$ [see the r.h.s. of Eq.~(\ref{ham})]. The functional derivatives acting on $\,\hat{S}^{(2n)}\,$ can be brought inside the trace and act directly on the Wilson lines $V$ and $V^\dagger$ in the following way:
\bea
\frac{\delta}{\delta \alpha ^a(x)} V_{x_i}^\dagger &=& i g \, \delta ^2(x_i-x) \, t^a \, V_{x_i}^\dagger \nn
\frac{\delta}{\delta \alpha ^a(x)} V_{x_i} &=& - \, i g \, \delta ^2(x_i-x) \, V_{x_i} \, t^a.
\label{deriv_wil}
\eea

The next step is to contract the indices $a$ and $b$ [see the r.h.s. of Eq.~(\ref{ham})]. The first term inside the parentheses of Eq.~(\ref{ham}) is the identity matrix, $1^{ab} = \delta^{ab}$. The second, third and fourth terms inside the parentheses may have their indices contracted with the $t$ matrices just after a small manipulation. Since each of these terms are products of two Wilson lines in the adjoint representation, we need to decompose them before they can be brought inside the trace. For the second term inside the parentheses of Eq.~(\ref{ham}) we have
\be
( U_x^\dagger U_y )^{ab} = ( U_x^\dagger)^{ac} (U_y)^{cb} \, .
\label{decomp_xy}
\ee
The elements $( U_x^\dagger)^{ac}$ and $(U_y)^{cb}$ are mere numbers, and can be moved freely inside the trace to contract with the $t^a$ and $t^b$ matrices through the relations
\bea
( U_x^\dagger)^{ac} \, t^a &=&V_x^\dagger \, t^c \, V_x \nn
               (U_y)^{cb} \, t^b &=& V_y^\dagger \, t^c \, V_y.
\label{contr_2}
\eea
Similar manipulations are done to contract the indices of the third and fourth terms inside the parentheses of Eq.~(\ref{ham}) with $t^a$ and $t^b$ matrices.

The third step is to integrate the r.h.s. of Eq.~(\ref{ham}) with respect to $x$ and $y$ in order to eliminate the delta functions originated after applying the functional derivatives [see Eq.~(\ref{deriv_wil})]. Since the Wilson lines do not depend on $\,x\,$ nor on $\,y\,$, the integration in these variables will only change the kernel. 

The last step is to use the Fierz identity to manipulate the traces and eliminate the $t$ matrices as follows:
\bea
Tr[t^cB \, t^cCA] \, = \, \frac{1}{2}Tr[B] \, Tr[AC] - \frac{1}{2N_c}Tr[ABC].
\label{fierz_2}
\eea

Our final result can be written as 
\bea
H\,\hat{S}^{(2n)} = \frac{\bar{\alpha}}{4 \pi} \int _z && \Big( \,\, H_1\,\hat{S}^{(2n)} \, + \, H_2\,\hat{S}^{(2n)}\nn 
&-& H_3\,\hat{S}^{(2n)} \, - \, H_4\,\hat{S}^{(2n)} \,\, \Big),
\label{hs2n}
\eea
where $\,\hat{S}^{(2n)}\,$ is defined in Eq.~(\ref{2n_op}), with $\,\bar{\alpha} = \alpha_s N_c/\pi \,$ and $\,\alpha_s = g^2/4\pi \,$. The explicit expressions for the action of $\{H_1, H_2, H_3, H_4\}$ on $\hat{S}^{(2n)}$ on the right hand side of Eq.~(\ref{hs2n}), are given in the Appendix. We have checked that our result is finite in the limit $z \rightarrow x_i$ for any external coordinate $x_i$. It is worth noting that all $N_c$ suppressed terms in Eq.~(\ref{hs2n}) cancel. We also make available for download a {\it Mathematica}~\cite{math} program which returns this expression after an integer value for $n$ is inserted~\cite{program}. 

\section{The dilute regime: $2n$ Reggeized gluon exchange}\label{III}

In this paper our goal is to establish a formal equivalence between the JIMWLK evolution equation in the dilute regime and the BJKP equation. Therefore we need to write Eq.~(\ref{ham}) in the dilute regime, i.e., when non-linear terms in the equation are ignored. This is most conveniently done by switching from the $S$ to $T$-matrices, defined by $\,\hat{S} = \,1-\hat{T}\,$. This substitution makes it easier to distinguish the linear terms from the non-linear ones since the linear regime is defined as $\hat{T} \ll 1$.  After switching from $\hat{S}$ to $1-\hat{T}$ the terms quadratic in $\hat{S}$ will become: $\,\hat{S}^q\,\hat{S}^p\,\to \, (1-\hat{T}^q)(1-\hat{T}^p)=1-\hat{T}^q-\hat{T}^p+\hat{T}^q\hat{T}^p$. We can then disregard the quadratic terms $\hat{T}^q\hat{T}^p$ because these are taken as negligible in the linear regime. One can also verify that all kernels multiplying $1$ will add up to zero. Equation (\ref{hs2n}) then becomes
\bea
H\,\hat{T}^{(2n)} = \frac{\bar{\alpha}}{4 \pi} \,\, \int _z && \Big( \,\, H_1\,\hat{T}^{(2n)} \, + \, H_2\,\hat{T}^{(2n)}\nn
&-& \, H_3\,\hat{T}^{(2n)} \, - \, H_4\,\hat{T}^{(2n)}\Big).
\label{ht2n}
\eea

To proceed further we expand each Wilson line in the $\hat{T}$ matrix to 1st order in the gluon field $\,\alpha\,$ and keep terms of order $\,O(\alpha^{2n})\,$ for $\,\hat{T}\,$ in both sides of Eq.~(\ref{ht2n}). The expansion of the Wilson lines is given by:
\bea
V_{x_i}& = &1 \, - \, ig\alpha _{x_i} \, + \, \cdots 
\nn
V_{x_i}^\dagger& = & 1 \, + \, ig\alpha _{x_i} \, + \, \cdots
\label{exp_wl}
\eea
Keeping only 1st order terms in $\,\alpha\,$ the expanded $\,\hat{T}\,$ operator is defined as
\bea
\hat{T}^{(2n)}_{\left(\prod _{k=1}^{2n} x_k\right)} \,\, \simeq \,\, g^{2n} \, \frac{1}{N_c} \, Tr[\alpha_{x_1}\alpha_{x_2} \cdots \alpha_{x_{2n-1}}\alpha_{x_{2n}}].
\label{weak_defin}
\eea
There is another contribution of order $O(\alpha^{2n})$ that we need to consider, namely those terms coming from taking 
 $\alpha_z^2$ terms in the expansion of the Wilson line $V_z$ (or $V_z^\dagger$). In this case we have to set one of the remaining Wilson lines equal to $1$ ($V_{x_i}\approx 1$) and take the 1st order term in the $\alpha$ expansion for the remaining Wilson lines. All other combinations either vanish or lead to external gluon fields with zero transverse momenta which are disregarded.

We now Fourier transform our equation for $\,\hat{T}\,$'s with $\,2n\,$ components to momentum space. The Fourier transform of the operator $\,\, \hat{T}^{(2n)}_{\left(\prod _{k=1}^{2n} x_k\right)} \,\,$ to momentum space is defined as
\bea
\tilde{T}^{(2n)}_{\left(\prod _{k=1}^{2n} l_k\right)} =  \int \, \left[ \, \prod _{k =1}^{2n} d^2x_{k} \, \right] \, e^{i \left( \sum _{k=1}^{2n} l_k \cdot x_k \right) } \hat{T}^{(2n)}_{\left(\prod _{k=1}^{2n} x_k\right)}\nn
\label{ftd}
\eea
where $l_k$ is the external momentum conjugate to the coordinate $x_k$. The final expression for the Fourier transform of the equation for $\,\hat{T}^{(2n)}\,$ to momentum space is
\begin{widetext}
\bea
\frac{d}{dY} \tilde{T}^{(2n)}_{\left(\prod _{k=1}^{2n} l_k\right)}&=&
 - \sum _{j = 1}^{2n} \, \left\lbrace \,\, \frac{\bar{\alpha} }{4\pi}  \, \int \,\, d^2p_t \, \left( \frac{l_j^2}{p_t^2(p_t+l_j)^2} \right) \,\,
\tilde{T}^{(2n)}_{\left(\prod _{k=1}^{2n} l_k\right)} 
\,\, \right\rbrace\nn
&+&\frac{\bar{\alpha}}{2\pi} \,\,
\int \, d^2p_t \,  \left( 
\frac{1}{p_t^2} \, + \, \frac{p_t\cdot l_{1}}{p_t^2l_{1}^2} \, - \, \frac{p_t\cdot l_{2n}}{p_t^2l_{2n}^2} \,-\, \frac{l_1\cdot l_{2n}}{l_1^2l_{2n}^2}
\right) \,\, \tilde{T}^{(2n)}_{(l_1+p_t)\, \left(\prod _{k=2}^{2n-1} l_k\right)\, (l_{2n}-p_t) }\nn
&+&\sum _{j = 2}^{2n} \, \left\lbrace \,\, \frac{\bar{\alpha}}{2\pi}  \, \, \int \, d^2p_t \, \left( 
\frac{1}{p_t^2} \, + \, \frac{p_t \cdot l_{j-1}}{p_t^2 \, l_{j-1}^2} \, - \, \frac{p_t \cdot l_j}{p_t^2 \, l_j^2} \,-\, \frac{l_{j-1}\cdot l_{j}}{l_{j-1}^2l_{j}^2}
\right) \,\,\,
\tilde{T}^{(2n)}_{\left(\prod _{k=1}^{j-2} l_k\right) (l_{j-1}+p_t) (l_{j}-p_t)  \left(\prod _{k=j+1}^{2n} l_k\right)} 
\,\, \right\rbrace.\nn
\label{ftfinal}
\eea
\end{widetext}
The last term inside parenthesis in the 2nd and 3rd lines of Eq.~(\ref{ftfinal}), comes from taking the 2nd order in the $\alpha_z$ expansion. We emphasize that $\,\tilde{T}\,$ above represents the expanded (in powers of the gluon field $\alpha$) Wilson lines. Finally we need to rewrite the operators $\,\tilde{T}^{(2n)}\,$ in terms of the charge density $\,\rho\,$ rather than the gauge field $\,\alpha\,$. For this we need to rewrite the kernel of the 3rd line of Eq.~(\ref{ftfinal}) as
\bea
&&\left[ \frac{1}{p_t^2} + \frac{p_t \cdot l_{j-1}}{p_t^2l_{j-1}^2} - \frac{p_t \cdot l_j}{p_t^2l_j^2} - \frac{l_{j-1} \cdot l_j}{l_{j-1}^2l_j^2} \right] =\nn
&&\frac{1}{2} \left[ \frac{(p_t+l_{j-1})^2}{p_t^2l_{j-1}^2} + \frac{(p_t-l_j)^2}{p_t^2l_j^2} - \frac{(l_{j-1}+l_j)^2}{l_{j-1}^2l_j^2} \right].
\eea
A similar expression is used to rewrite the kernel of the 2nd line of Eq.~(\ref{ftfinal}).
We now use the relation between the gauge field $\,\alpha\,$ and the color charge density $\,\rho\,$
\bea
\alpha (p_t) \sim \frac{\rho (p_t)}{p_t^2}.
\label{relalfrho}
\eea
We also define the operator $T^{(2n)}$, which depends on $\,\rho\,$ rather than $\,\alpha\,$
\bea
T^{(2n)}_{\left(\prod _{k=1}^{2n} l_k\right)} \,\, = \,\, \frac{1}{N_c} Tr \left[ \,\prod _{k=1}^{2n} \rho(l_k) \,\right]
\label{deft2nrho}
\eea
Then, by multiplying both sides of Eq. (\ref{ftfinal}) with $\,\left(\prod _{k=1}^{2n} l_k^2\right)\,$ and using the relation given in Eq.~(\ref{relalfrho}), we make some final algebraic manipulations and obtain
\begin{widetext}
\bea
&&\frac{d}{dY} T^{(2n)}_{\left(\prod _{k=1}^{2n} l_k\right)} =
- \frac{\bar{\alpha} }{2\pi}  \, \sum _{j = 1}^{2n} \,
\int \,\, d^2p_t \, \left[ \frac{l_j^2}{p_t^2 \, [p_t^2 + (p_t - l_j)^2]} \right] \,\,
T^{(2n)}_{\left(\prod _{k=1}^{2n} l_k\right)} \nn
&+& \, \frac{\bar{\alpha}}{4\pi} \,\,
\int \, d^2p_t \,  
\left[ \frac{l_1^2}{p_t^2\, (p_t + l_1)^2} + \frac{l_{2n}^2}{p_t^2\, (p_t - l_{2n})^2} - 
\frac{(l_1 + l_{2n})^2}{(p_t + l_1)^2 \, (p_t - l_{2n})^2} \right]
\,\, T^{(2n)}_{(l_1+p_t)\, \left(\prod _{k=2}^{2n-1} l_k\right)\, (l_{2n}-p_t) } \nn
&+&  \frac{\bar{\alpha}}{4\pi} \, 
\sum _{j = 2}^{2n} \, \, \int \, d^2p_t \, 
\left[ \frac{l_{j-1}^2}{p_t^2\, (p_t + l_{j-1})^2} + \frac{l_j^2}{p_t^2\, (p_t - l_j)^2} - 
\frac{(l_{j-1} + l_j)^2}{(p_t + l_{j-1})^2 \, (p_t - l_j)^2} \right]
\,\,\,
T^{(2n)}_{\left(\prod _{k=1}^{j-2} l_k\right) (l_{j-1}+p_t) (l_{j}-p_t)  \left(\prod _{k=j+1}^{2n} l_k\right)}.\nn
\label{t2n_rho_final}
\eea
\end{widetext}
Equation~(\ref{t2n_rho_final}) is our final result. We have checked that it is infrared finite ($p_t \rightarrow 0$) due to cancellation of infrared divergent terms between the virtual (first line) and real (second and third lines) corrections. It reduces to the previously known results for $n=2, 3$ and $ 4$ \cite{cgc-odd, cgc-quad}. This formally establishes the equivalence of the JIMWLK equation for $x$ evolution of the correlator of $2 n$ Wilson lines in the dilute regime with the BJKP equation for the evolution of $2 n$ Reggeized gluon state~\cite{bjkp}. 

\section{Summary and Conclusions}\label{concl}

In this work we have shown the equivalence of the JIMWLK equation for the $x$ evolution of the $2 n$ Wilson line correlator in the dilute regime with the BJKP equation for the evolution of $2 n$ Reggeized gluon state. While our result is new in the sense that it explicitly shows the equivalence of the two approaches, it is not unexpected. The equivalence of the two approaches has already been established for the $3$ and $4$ Reggeized gluon exchange~\cite{cgc-odd, cgc-quad}. The more interesting and useful result is that the expansion of Wilson lines to a given order in the gluon field is the systematic way of recovering the physics of BJKP equation, and it suggests that one can use this expansion to derive other more interesting results. For example, the expansion of Wilson lines in powers of the gluon fields has already been used to derive the $3$-pomeron vertex ($2$ Reggeized gluons going into $4$ Reggeized gluons) from JIMWLK equation~\cite{3pom}. We expect that similar expressions can be derived explicitly for all other $n \rightarrow m$ vertices. Work in this direction is in progress and will be reported elsewhere. Another aspect which would be interesting to pursue is to derive NLO corrections to the BJKP equation in the CGC formalism. This is now possible due to the recent computation of NLO corrections to the JIMWLK equation~\cite{nlo-jimwlk}. We plan to take this up in the near future.

\section*{Acknowledgments}
E.R.C. gratefully acknowledges the Brazilian Funding Agency FAPESP for financial support (contract: 2013/21759-5). Support for this work has been received in part from DGAPA-UNAM under grant number PAPIIT-IN103811 and CONACyT-M\'exico under grant number 128534. J.J-M. acknowledges support by the DOE Office of Nuclear Physics through Grant No.\ DE-FG02-09ER41620 and
from The City University of New York through the PSC-CUNY Research Award Program, grant 67732-0045. A.A., L.A.H. and M.E.T. gratefully acknowledge the hospitality of Baruch College at CUNY, where the early stages of this work was done.

\section*{Appendix}

Here we give the explicit expressions for the individual contributions in the right hand side of Eq.~(\ref{hs2n})
\begin{widetext}
\bea
H_1\,\hat{S}^{(2n)} &=& - \sum _{i=1}^{n} \,\, \big[ M_{x_{2i-1}x_{2i}z}\big] \,\, 
\hat{S}^{(2n)}_{\left(\prod _{k=1}^{2n} x_k\right)} 
\nonumber \\
&+&\sum _{j = 2 \, ; \, j\, >\, i}^n \,\, \sum _{i = 1}^{n-1} \,\,  \big[\, \, M_{x_{2i-1}x_{2j-1}z} \, - \, M_{x_{2i-1}x_{2j}z} \, - \, M_{x_{2i}x_{2j-1}z} \, + \, M_{x_{2i}x_{2j}z} \, \, \big] \,  \hat{S}^{(2j-2i)}_{\left(\prod _{k=2i}^{2j-1} x_k\right)}\hat{S}^{(2n-2j+2i)}_{\left(\prod _{k=1}^{2i-1} x_k \right) \left(\prod _{k=2j}^{2n} x_k\right)}
\nonumber \\
\label{h1s2n}
\eea

\bea
H_2\,\hat{S}^{(2n)} &=& - \sum _{i=1}^{n} \,\, \left[\, M_{x_{2i-1}x_{2i}z} \, \right] \,\, \hat{S}^{(2)}_{\left(x_{2i-1}x_{2i}\right)}\hat{S}^{(2n-2)}_{\left(\prod _{k=1}^{2i-2} x_k\right) \left(\prod _{k=2i+1}^{2n} x_k\right)}  
\nonumber \\
&& + \,\, \sum _{j = 2 \, ; \, j\, >\, i}^n \,\, \sum _{i = 1}^{n-1} \big[\, M_{x_{2i-1}x_{2j-1}z} \, \big] \, \hat{S}^{(2j-2i)}_{\left(\prod _{k=2i-1}^{2j-2} x_k\right)}\hat{S}^{(2n-2j+2i)}_{\left(\prod _{k=1}^{2i-2} x_k\right)\left(\prod _{k=2j-1}^{2n} x_k\right)} 
\nonumber \\
&& - \,\, \sum _{j = 2 \, ; \, j\, >\, i}^n \,\, \sum _{i = 1}^{n-1} \big[\, M_{x_{2i-1}x_{2j}z} \, \big] \,\, \hat{S}^{(2j-2i+2)}_{\left(\prod _{k=2i-1}^{2j} x_k\right)}\hat{S}^{(2n-2j+2i-2)}_{\left(\prod _{k=1}^{2i-2} x_k\right)\left(\prod _{k=2j+1}^{2n} x_k\right)}  
\nonumber \\
&& - \,\, \sum _{j = 2 \, ; \, j\, >\, i}^n \,\, \sum _{i = 1}^{n-1} \big[\, M_{x_{2i}x_{2j-1}z} \, \big] \,\, \hat{S}^{(2j-2i-2)}_{\left(\prod _{k=2i+1}^{2j-2} x_k\right)}\hat{S}^{(2n-2j+2i+2)}_{\left(\prod _{k=1}^{2i} x_k\right)\left(\prod _{k=2j-1}^{2n} x_k\right)} 
\nonumber \\
&& + \,\, \sum _{j = 2 \, ; \, j\, >\, i}^n \,\, \sum _{i = 1}^{n-1} \big[\, M_{x_{2i}x_{2j}z} \, \big] \,\, \hat{S}^{(2j-2i)}_{\left(\prod _{k=2i+1}^{2j} x_k\right)}\hat{S}^{(2n-2j+2i)}_{\left(\prod _{k=1}^{2i} x_k\right)\left(\prod _{k=2j+1}^{2n} x_k\right)}
\label{h2s2n}
\eea

\bea
H_3\,\hat{S}^{(2n)}&=& - \sum _{i=1}^{n} \,\, \left[\, M_{x_{2i-1}x_{2i}z} \, \right] \,\, \frac{1}{2}\hat{S}^{(2)}_{z \, x_{2i-1}}\hat{S}^{(2n)}_{\left(\prod _{k=1}^{2i-2} x_k\right) \,z\, \left( \prod _{k=2i}^{2n} x_k\right)} 
\nonumber \\
&& - \, \sum _{i=1}^{n} \,\, \left[\, M_{x_{2i-1}x_{2i}z} \, \right] \,\, \frac{1}{2}\hat{S}^{(2)}_{z\, x_{2i}}\hat{S}^{(2n)}_{\left( \prod _{k=1}^{2i-1} x_k  \right) \,z\, \left(\prod _{k=2i+1}^{2n} x_k \right)} 
\nonumber \\
&& + \, \sum _{j = 2 \, ; \, j\, >\, i}^n \,\, \sum _{i = 1}^{n-1} \big[\, M_{x_{2i-1}x_{2j-1}z} \, - \, M_{x_{2i-1}x_{2j}z} \, \big]  \,\, \frac{1}{2}\hat{S}^{(2j-2i+2)}_{z\, \left(\prod _{k=2i-1}^{2j-1} x_k \right)}\hat{S}^{(2n-2j+2i)}_{\left(\prod _{k=1}^{2i-2} x_k\right) \,z\, \left(\prod _{k=2j}^{2n} x_k\right)} 
\nonumber \\
&& + \,\, \sum _{j = 2 \, ; \, j\, >\, i}^n \,\, \sum _{i = 1}^{n-1} \big[\, - \, M_{x_{2i}x_{2j-1}z} \, + \, M_{x_{2i}x_{2j}z} \, \big] \,\, \frac{1}{2}\hat{S}^{(2j-2i)}_{z\, \left( \prod _{k=2i+1}^{2j-1} x_k \right)}\hat{S}^{(2n-2j+2i+2)}_{\left(\prod _{k=1}^{2i} x_k\right) \,z\, \left( \prod _{k=2j}^{2n} x_k \right)} 
\nonumber \\
&& + \, \sum _{j = 2 \, ; \, j\, >\, i}^n \,\, \sum _{i = 1}^{n-1} \big[\, M_{x_{2j-1}x_{2i-1}z} \, - \, M_{x_{2j-1}x_{2i}z} \, \, \big] \,\, \frac{1}{2}\hat{S}^{(2j-2i)}_{z\,\left( \prod _{k=2i}^{2j-2} x_k \right)}\hat{S}^{(2n-2j+2i+2)}_{\left( \prod _{k=1}^{2i-1} x_k  \right) \,z\, \left(\prod _{k=2j-1}^{2n} x_k \right)} \nonumber \\
&& + \,\, \sum _{j = 2 \, ; \, j\, >\, i}^n \,\, \sum _{i = 1}^{n-1} \big[\, - \, M_{x_{2j}x_{2i-1}z} \, + \, M_{x_{2j}x_{2i}z} \, \, \big] \,\, \frac{1}{2}\hat{S}^{(2j-2i+2)}_{z\, \left( \prod _{k=2i}^{2j} x_k \right)}\hat{S}^{(2n-2j+2i)}_{\left( \prod _{k=1}^{2i-1} x_k \right) \,z\, \left(\prod _{k=2j+1}^{2n} x_k \right)}
\label{h3s2n}
\eea

\bea
H_4\,\hat{S}^{(2n)}&=& - \sum _{i=1}^{n} \,\, \left[\, M_{x_{2i-1}x_{2i}z} \, \right] \,\, \frac{1}{2}\hat{S}^{(2)}_{z \, x_{2i}}\hat{S}^{(2n)}_{\left( \prod _{k=1}^{2i-1} x_k  \right) \,z\, \left( \prod _{k=2i+1}^{2n} x_k \right)} 
\nonumber \\
&& - \, \sum _{i=1}^{n} \,\, \left[\, M_{x_{2i-1}x_{2i}z} \, \right] \,\, \frac{1}{2}\hat{S}^{(2)}_{z\, x_{2i-1}}\hat{S}^{(2n)}_{\left(\prod _{k=1}^{2i-2} x_k\right) \,z\, \left(\prod _{k=2i}^{2n} x_k\right)} 
\nonumber \\
&& + \, \sum _{j = 2 \, ; \, j\, >\, i}^n \,\, \sum _{i = 1}^{n-1} \big[\, M_{x_{2i-1}x_{2j-1}z} \, \, - \, M_{x_{2i}x_{2j-1}z} \, \big]  \,\, \frac{1}{2}\hat{S}^{(2j-2i)}_{z\, \left( \prod _{k=2i}^{2j-2} x_k \right)}\hat{S}^{(2n-2j+2i+2)}_{\left( \prod _{k=1}^{2i-1} x_k  \right) \,z\, \left( \prod _{k=2j-1}^{2n} x_k \right)} \nonumber \\
&& + \,\, \sum _{j = 2 \, ; \, j\, >\, i}^n \,\, \sum _{i = 1}^{n-1} \big[\, - \, M_{x_{2i-1}x_{2j}z} \, + \, M_{x_{2i}x_{2j}z} \, \big] \,\, \frac{1}{2}\hat{S}^{(2j-2i+2)}_{z\, \left( \prod _{k=2i}^{2j} x_k \right)}\hat{S}^{(2n-2j+2i)}_{\left( \prod _{k=1}^{2i-1} x_k  \right) \,z\, \left( \prod _{k=2j+1}^{2n} x_k \right)}
\nonumber \\
&& + \, \sum _{j = 2 \, ; \, j\, >\, i}^n \,\, \sum _{i = 1}^{n-1} \big[\, M_{x_{2j-1}x_{2i-1}z} \, - \, M_{x_{2j}x_{2i-1}z} \, \big] \,\, \frac{1}{2}\hat{S}^{(2j-2i+2)}_{z\, \left( \prod _{k=2i-1}^{2j-1} x_k \right)}\hat{S}^{(2n-2j+2i)}_{\left( \prod _{k=1}^{2i-2} x_k \right) \,z\, \left( \prod _{k=2j}^{2n} x_k \right)} 
\nonumber \\
&& + \,\, \sum _{j = 2 \, ; \, j\, >\, i}^n \,\, \sum _{i = 1}^{n-1} \big[\, - \, M_{x_{2j-1}x_{2i}z} \, + \, M_{x_{2j}x_{2i}z} \, \big] \,\, \frac{1}{2}\hat{S}^{(2j-2i)}_{z\, \left( \prod _{k=2i+1}^{2j-1} x_k \right)}\hat{S}^{(2n-2j+2i+2)}_{\left( \prod _{k=1}^{2i} x_k \right) \,z\, \left( \prod _{k=2j}^{2n} x_k \right)}.
\label{h4s2n}
\eea
\end{widetext}
Note that the summation in $\,j\,$ is always restricted by the condition $\,j>i\,$. Moreover, as described above, the upper index attached to $\,\hat{S}\,$ represents the number of Wilson lines that make up this operator, and the lower index represents the coordinate dependence of these Wilson lines. For example, the operator $\,\hat{S}^{(2n-2j+2i+2)}_{\left( \prod _{k=1}^{2i} x_k \right) \,z\, \left( \prod _{k=2j}^{2n} x_k \right)}\,$ is given by the normalized trace of ``$\,2n-2j+2i+2\,$'' Wilson lines. From these, the first ``$\,2i\,$'' Wilson lines have coordinates $\,x_1\,x_2\,\cdots\,x_{2i}\,$, being followed by one Wilson line with coordinate $\,z\,$ and then by ``$\,2n-2j+1\,$'' Wilson lines with coordinates $\,\,x_{2j}\,\,x_{2j+1}\,\,\cdots\,\,x_{2n}\,\,$.

We make available for download a {\it Mathematica}~\cite{math} program which returns the right hand side of eq. (\ref{hs2n}) after an integer value for $n$ is inserted~\cite{program}.


\begin{thebibliography}{99}
\bibitem{program}
The code may be downloaded from 

{\tt faculty.baruch.cuny.edu/naturalscience/physics/}
{\tt Jalilian-Marian/}

or from 

{\tt paginas.fisica.uson.mx/elena.tejeda/code.nb}
\end{thebibliography}
\end{document}